\documentclass[journal,comsoc,twoside]{IEEEtran}
  
    \usepackage{newtxtext,newtxmath}
\ifCLASSOPTIONcompsoc
  \usepackage[caption=false,font=footnotesize,labelfont=sf,textfont=sf]{subfig}
\else
  \usepackage[caption=false,font=footnotesize,labelfont=small]{subfig}
\fi
\ifCLASSOPTIONcaptionsoff
  \usepackage[nomarkers]{endfloat}
 \let\MYoriglatexcaption\caption
 \renewcommand{\caption}[2][\relax]{\MYoriglatexcaption[#2]{#2}}
\fi

\usepackage[utf8]{inputenc}
\usepackage{ragged2e}
\usepackage{units}
\usepackage{graphicx}
\usepackage{tabulary}
\usepackage{booktabs}
\usepackage{mathtools}  

\usepackage{amsmath} 
\usepackage{amsthm} 

\usepackage{amssymb}
\usepackage{amsfonts}
\usepackage[noadjust]{cite}
\usepackage{balance} 
\usepackage{tabularx}
\usepackage{comment}    
\usepackage{setspace}
\usepackage{xspace}
\usepackage{acronym}
\usepackage{multirow}
\usepackage{upgreek} 
\usepackage{nicefrac} 
\usepackage{ifthen}
\usepackage{csvsimple}
\usepackage{hyperref}
\usepackage{flexisym} 
\usepackage{fp} 
\usepackage{caption}
\usepackage{titlecaps}
\usepackage{optidef}
\usepackage[thinc]{esdiff} 
\theoremstyle{definition}

\usepackage{xargs} 
\usepackage{xcolor}
\usepackage{letltxmacro}

\LetLtxMacro{\oldsqrt}{\sqrt}
\renewcommand{\sqrt}[2][\mkern8mu]{\mkern-4mu\mathop{\oldsqrt[#1]{#2}}}

\usepackage{titlecaps}

\usepackage{graphicx}
\usepackage{flushend}

    \usepackage{acronym}
\acrodef{eMBB}{Enhanced Mobile Broadband}
\acrodef{URLLC}{Ultra-Reliable and Low-Latency Communications}
\acrodef{FFT}{Fast Fourier Transform}
\acrodef{iFFT}{inverse FFT}
\acrodef{ARQ}{Automatic Repeat Request}

\acrodef{TWDMA}{Time and Wavelength Division Multiple Access}
\acrodef{CDF}{cumulative distribution function}
\acrodef{BA}{Bandwidth Allocation}
\acrodef{SBA}{Static Bandwidth Allocation}
\acrodef{BBU}{Base-Band Unit}
\acrodef{RRH}{Remote Radio Heads}
\acrodef{DRAN}{Distributed RANs}
\acrodef{5G}{Fifth Generation Mobile Networks}
\acrodef{B5G}{Beyond 5G}
\acrodef{CRAN}{Cloud Radio Access Network}
\acrodef{CPRI}{Common Public Radio Interface}
\acrodef{eCPRI}{enhanced Common Public Radio Interface}
\acrodef{MFH}{Mobile Fronthauling}
\acrodef{DU}{Distributed Unit}
\acrodef{CU}{Central Unit}
\acrodef{RU}{Radio Unit}
\acrodef{PRA}{Prioritized Random Access}
\acrodef{ACK}{acknowledgement}
\acrodef{3GPP}{3rd Generation Partnership Project}
\acrodef{ACB}{Access Class Barring}
\acrodef{BSR}{Buffer Status Report}
\acrodef{BQA}{Bandwidth and QoS Aware}
\acrodef{CN}{Core Network}
\acrodef{CBR}{constant bit rate}
\acrodef{eNodeB}{evolved NodeB}
\acrodef{eNB}{evolved NodeB}
\acrodef{E-UTRAN}{Evolved Universal Terrestrial Radio Access Network}
\acrodef{FTP}{File Transfer Protocol}
\acrodef{GBR}{Guaranteed Bit Rate}
\acrodef{H2H}{human-to-human}
\acrodef{HARQ}{Hybrid Automatic Repeat Request}
\acrodef{HTC}{Human-type Communication}
\acrodef{IoT}{internet of things}
\acrodef{LTE}{Long Term Evolution}
\acrodef{LTE-A}{LTE-Advanced}
\acrodef{M2M}{Machine-to-Machine}
\acrodef{MTC}{Machine-type Communications}
\acrodef{PDCCH}{Packet Downlink Control Channel}
\acrodef{ePDCCH}{enhanced \acs{PDCCH}}
\acrodef{PDSCH}{Physical Downlink Shared Channel}
\acrodef{PUSCH}{Packet Uplink Shared Channel}
\acrodef{PLR}{packet loss ratio}
\acrodef{PRACH}{Physical Random Access Channel}
\acrodef{PDB}{Packet Delay Budget}
\acrodef{QCI}{QoS Class Identifier}
\acrodef{QoS}{quality of service}
\acrodef{RA}{Random-access}
\acrodef{RACH}{Random Access Channel}
\acrodef{RAN}{Radio Access Network}
\acrodef{RAR}{Random Access Response}
\acrodef{ROI}{RAN Overload Indicator}
\acrodef{RRC}{Radio Resource Control}
\acrodef{RRS}{RACH Resource Separation}
\acrodef{SR}{Scheduling Request}
\acrodef{UE}{User Equipment}
\acrodef{ML}{Machine Learning}
\acrodef{UL}{Uplink}
\acrodef{TA}{Time Alignment}
\acrodef{PUCCH}{Physical Uplink Control Channel}
\acrodef{PRB}{Physical Resource Block}
\acrodef{PCFICH}{Physical Control Format Indicator Channel}
\acrodef{PHICH}{Physical HARQ Indicator Channel}
\acrodef{DCI}{Downlink Control Information}
\acrodef{CCE}{Control Channel Element}
\acrodef{OFDM}{Orthogonal Frequency Division Multiplexing}
\acrodef{CFI}{Control Format Indicator}
\acrodef{OFDMA}{Orthogonal Frequency Division Multiple Access}
\acrodef{SC-FDMA}{Single-Carrier Frequency Division Multiple Access}
\acrodef{MME}{Mobility Management Entity}
\acrodef{S-GW}{Serving Gateway}
\acrodef{P-GW}{Packet Data Network Gateway}
\acrodef{TTI}{Transmission Time Interval}
\acrodef{nGBR}{non-GBR}
\acrodef{SDF}{Service Data Flow}
\acrodef{TFT}{Traffic Flow Template}
\acrodef{CP}{cyclic prefix}
\acrodef{FDD}{Frequency Division Duplexing}
\acrodef{TDD}{Time Division Duplexing}
\acrodef{EPS}{Evolved Packet System}
\acrodef{MCS}{Modulation and Coding Scheme}
\acrodef{DCCH}{Downlink Control Channels}
\acrodef{EPC}{Evolved Packet Core}
\acrodef{PER}{Packet Error Rate}
\acrodef{EAB}{Extended Access Barring}
\acrodef{TDPS}{Time-domain Packet Scheduling}
\acrodef{DL}{Downlink}
\acrodef{DSCH}{Downlink Shared Channel}
\acrodef{BLER}{Block Error Rate}
\acrodef{MNO}{mobile network operator}
\acrodef{PS}{Packet Scheduling}
\acrodef{TD}{Time-domain}
\acrodef{FD}{Frequency-domain}
\acrodef{FDPS}{Frequency-domain Packet Scheduling}
\acrodef{QPSK}{quadrature phase-shift keying}
\acrodef{RRM}{Radio Resource Management}
\acrodef{RAC}{Radio Admission Control}
\acrodef{RB}{Resource Block}
\acrodef{VoIP}{Voice-over-IP}
\acrodef{UCCH}{Uplink Control Channel}
\acrodef{HoL}{Head of the Line}
\acrodef{RLC}{Radio Link Control}
\acrodef{AL}{Aggregation Level}
\acrodef{SRS}{Sounding Reference Signal}
\acrodef{SPS}{Semi-Persistent Scheduling}
\acrodef{TPC}{Transmit Power Control}

\acrodef{3GPP}{Third Generation Partnership Project}
\acrodef{ACB}{Access Class Barring}
\acrodef{BSR}{Buffer Status Report}
\acrodef{CN}{Core Network}
\acrodef{eNodeB}{evolved NodeB}
\acrodef{GBR}{Guarantee Bit Rate}
\acrodef{H2H}{human-to-human}
\acrodef{HARQ}{Hybrid Automatic Repeat Request}
\acrodef{HTC}{Human Type Communication}
\acrodef{LTE}{Long Term Evolution}
\acrodef{LTE-A}{LTE-Advanced}
\acrodef{M2M}{Machine-to-Machine}
\acrodef{MAC}{Medium Access Control}
\acrodef{MTC}{Machine-Type Communications}
\acrodef{mMTC}{Massive Machine-Type Communications}
\acrodef{PRACH}{Physical Random Access Channel}
\acrodef{PDB}{Packet Delay Budget}
\acrodef{QCI}{QoS Class Identifier}
\acrodef{QoS}{quality of service}
\acrodef{RA}{Random Access}
\acrodef{RACH}{Random Access Channel}
\acrodef{RAN}{Radio Access Network}
\acrodef{RAR}{Random Access Response}
\acrodef{RAP}{Random Access Priorized}
\acrodef{ROI}{RAN Overload Indicator}
\acrodef{RAO}{Random Access Opportunity}
\acrodef{RRC}{Radio Resource Control}
\acrodef{RRS}{RACH Resource Separation}
\acrodef{SR}{Scheduling Request}
\acrodef{UE}{User Equipment}
\acrodef{UL}{Uplink}
\acrodef{TA}{Time Alignment}
\acrodef{PPA}{Preamble-Priority Aware}
\acrodef{PPA+}[ePPA]{Enhanced Preamble-Priority-Aware}
\acrodef{CAM}{Critical Alarm Messages}
\acrodef{RAP}{RA-Priorized}
\acrodef{LTA}{Lifetime-Aware}
\acrodef{GBR-LTA}{GBR-Priorized LTA}
\acrodef{CSS}{common search space}
\acrodef{DSS}{dedicated search space}
\acrodef{MPDCCH}{MTC \acs{PDCCH}}
\acrodef{UL-SCH}{Uplink Share Channel}
\acrodef{CR}{Contention Resolution}
\acrodef{LTE-Sim}{LTE Simulator}
\acrodef{FTTH}{Fiber to the Home}

\acrodef{ATM}{asynchronous transfer mode}
\acrodef{GEM}{GPON encapsulation method}
\acrodef{DBA}{dynamic bandwidth allocation}
\acrodef{EPON}{Ethernet PON}
\acrodef{ONU}{optical network unit}
\acrodef{ODN}{optical distribution network}
\acrodef{PON}{passive optical network}
\acrodef{SLA}{service level agreement}
\acrodef{VoIP}{voice over IP}
\acrodef{GPON}{Gigabit-capable \ac{PON}}
\acrodef{IPACT}{interleaved polling with adaptive cycle time}
\acrodef{DiffServ}{differentiated services}
\acrodef{OLT}{optical line terminal}
\acrodef{IEEE}{Institute of Electrical and Electronics Engineers}
\acrodef{TDM}{time division multiplexing}
\acrodef{TDMA}{time division multiple access}
\acrodef{MPCP}{multipoint control protocol}
\acrodef{WDM}{wavelength division multiplexing}
\acrodef{TWDM}{time and wavelength division multiplexing}
\acrodef{WDMA}{wavelength division multiple access}
\acrodef{TWDMA}{time and wavelength division multiple access}
\acrodef{TWMD}{time/wavelength division multiplexing}
\acrodef{FTTx}{fiber to the x}
\acrodef{PHY}{physical layer}
\acrodef{MAC}{media access control layer}
\acrodef{CoS}{class of services}
\acrodef{gGIANT}{group-GIANT}
\acrodef{XGPON}{10-Gigabit-capable PON}
\acrodef{GIANT}{Giga-PON access network}
\acrodef{T-CONT}{transmission container}
\acrodef{gT-CONT}{grouped T-CONT}
\acrodef{gAllocId}{group allocation ID}
\acrodef{XGEM}{GPON encapsulation mode}
\acrodef{EF}{expedited forwarding}
\acrodef{BE}{best effort}
\acrodef{AF}{assured forwarding}
\acrodef{BS}{Bandwidth Slicing}
\acrodef{mBS}{modified-BS}
\acrodef{IP}{interleaved polling}
\acrodef{DWBA}{dynamic wavelength and bandwidth allocation}
\acrodef{InP}{infrastructure provider}
\acrodef{TCO}{total cost of ownership}
\acrodef{CAPEX}{CAPital EXpenditure}
\acrodef{OPEX}{OPerating EXpenditure}
\acrodef{VNO}{virtual network operator}
\acrodef{MOS-IPACT}{IPACT with multi-ONU SLAs support}
\acrodef{SNS}{SubNetwork Slice}
\acrodef{MAC}{medium access control}
\acrodef{IP}{interleaved polling}
\acrodef{OLS}{onu load status}
\acrodef{RTT}{round to trip}
\acrodef{10G-EPON}{10 Gbit/s ethernet passive optical network}
\acrodef{X-GPON}{10 Gbit/s GPON}
\acrodef{ITU}{International Telecommunication Union}
\acrodef{DES}{Delayed Excess Scheduling}
\acrodef{GSF}{Grant scheduling frameworks}
\acrodef{GSP}{Grant sizing policy}
\acrodef{PEDB}{Policies for excess bandwidth distribution}
\acrodef{THR}{throughput}
\acrodef{FANS}{fixed access network sharing}
\acrodef{FE}{fair excess}
\acrodef{EC}{excess control}
\acrodef{HPS}{High Priority Subgroup First}
\acrodef{XGTC}{XG-PON transmission convergence}
\acrodef{NO}{network operator}

\acrodef{GSF}{grant scheduling framework}
\acrodef{GWP}{grant windows-size policy}
\acrodef{EDP}{excess distribution policy}
\acrodef{GSP}{grant scheduling policy}
\acrodef{TSF}{thread scheduling framework}
\acrodef{RTT}{round trip time}
\acrodef{InP}{Infrastructure service provider}
\acrodef{FL}{Federated Learning}
\acrodef{10G-EPON}{10 Gbit/s Ethernet Passive Optical Network}
\acrodef{3GPP}{3rd Generation Partnership Project}
\acrodef{3GPP}{Third Generation Partnership Project}
\acrodef{BAM}{Bandwidth-slicing with Adaptive-cycle and Multiple-wavelengths algorithm}
\acrodef{CNN}{Convolutional Neural Network}
\acrodef{ACB}{Access Class Barring}
\acrodef{ACK}{Acknowledgement}
\acrodef{AF}{Assured Forwarding}
\acrodef{Alloc-ID}{Allocation Identify}
\acrodef{APON}{ATM PON}
\acrodef{ASIC}{Application Specific Integrated Circuit}
\acrodef{ATM}{Asynchronous Transfer Mode}
\acrodef{AL}{Aggregation Level}

\acrodef{BE}{Best Effort}
\acrodef{BQA}{Bandwidth and QoS Aware}
\acrodef{BGP}{Bandwidth Guaranteed Polling}
\acrodef{BW map}{Bandwidth Mapping}
\acrodef{BLER}{Block Error Rate}
\acrodef{BSR}{Buffer Status Report}
\acrodef{BPON}{Broadband PON}

\acrodef{CAPEX}{Capital Expenditure}
\acrodef{CO}{Central Office}
\acrodef{CBR}{Constant Bit Rate}
\acrodef{CCE}{Control Channel Element}
\acrodef{CFI}{Control Format Indicator}
\acrodef{CN}{Core Network}
\acrodef{CR}{Contention Resolution}
\acrodef{CoS}{Class of Services}
\acrodef{CAM}{Critical Alarm Messages}
\acrodef{CP}{Cyclic Prefix}
\acrodef{CSS}{Common Search Space}

\acrodef{DBA}{Dynamic Bandwidth Allocation}
\acrodef{DBAM}{Dynamic Bandwidth Allocation with Multiple Services}
\acrodef{DSCH}{Downlink Shared Channel}
\acrodef{DCCH}{Downlink Control Channels}
\acrodef{DCI}{Downlink Control Information}
\acrodef{DES}{Delayed Excess Scheduling}
\acrodef{DSS}{Dedicated Search Space}
\acrodef{DiffServ}{Differentiated Services}
\acrodef{DL}{Downlink}
\acrodef{DPOA}{Dynamic Polling Order Arrangement}
\acrodef{DWBA}{Dynamic Wavelength and Bandwidth Allocation}

\acrodef{EC}{Excess Control}
\acrodef{EDP}{Excess Distribution Policy}
\acrodef{EF}{Expedited Forwarding}
\acrodef{EAB}{Extended Access Barring}
\acrodef{ePDCCH}{Enhanced \acs{PDCCH}}
\acrodef{eNodeB}{Evolved NodeB}
\acrodef{eNB}{Evolved NodeB}
\acrodef{E-UTRAN}{Evolved Universal Terrestrial Radio Access Network}
\acrodef{EPS}{Evolved Packet System}
\acrodef{EPC}{Evolved Packet Core}
\acrodef{EPDF}{Empirical Probability Density Function}
\acrodef{EPON}{Ethernet PON}

\acrodef{FE}{Fair Excess}
\acrodef{FPGA}{Field-programmable Gate Array}
\acrodef{FSD-SLA}{Fair Sharing with Dual SLAs}
\acrodef{FTP}{File Transfer Protocol}
\acrodef{FTTb}{Fiber To The Business}
\acrodef{FTTB}{Fiber To The Building}
\acrodef{FTTc}{Fiber To The cell}
\acrodef{FTTC}{Fiber To The Curb}
\acrodef{FTTO}{Fiber To The Office}
\acrodef{FTTH}{Fiber To The Home}
\acrodef{FTTdp}{Fiber To The Distribution Point}
\acrodef{FTTN}{Fiber To The Node}
\acrodef{FTTP}{Fiber To The Premises}
\acrodef{FTTx}{Fiber To The x}
\acrodef{FANS}{Fixed Access Network Sharing}
\acrodef{FD}{Frequency-domain}
\acrodef{FDPS}{Frequency-domain Packet Scheduling}
\acrodef{FDD}{Frequency Division Duplexing}

\acrodef{GEM}{GPON Encapsulation Method}
\acrodef{GIANT}{Giga-PON Access Network}
\acrodef{gAllocId}{Group Allocation ID}
\acrodef{gGIANT}{Group-GIANT}
\acrodef{GPON}{Gigabit Capable \ac{PON}}
\acrodef{GSF}{Grant Scheduling Framework}
\acrodef{GSP}{Grant Scheduling Policy}
\acrodef{gT-CONT}{Grouped T-CONT}
\acrodef{GTC}{GPON Transmission Convergence}
\acrodef{GWP}{Grant Windows-sizing Policy}

\acrodef{HGP}{Hybrid Granting Protocol}
\acrodef{HoL}{Head of the Line}
\acrodef{HPP}{High Priority Packets First}
\acrodef{HPS}{High Priority Subgroup First}
\acrodef{H2H}{Human-to-Human}
\acrodef{HTC}{Human Type Communication}
\acrodef{HARQ}{Hybrid Automatic Repeat Request}

\acrodef{IC}{Computing Institute}
\acrodef{IEEE}{Institute of Electrical and Electronics Engineers}
\acrodef{InP}{Infrastructure Provider}
\acrodef{IoT}{Internet of Things}
\acrodef{IP}{Interleaved Polling}
\acrodef{IPACT}{Interleaved Polling with Adaptive Cycle Time}
\acrodef{ITU}{International Telecommunication Union}

\acrodef{LLID}{Logical Link ID}
\acrodef{LTA}{Lifetime-Aware}
\acrodef{LTE}{Long Term Evolution}
\acrodef{LTE-A}{LTE-Advanced}
\acrodef{LTE-Sim}{LTE Simulator}
\acrodef{LNF}{Largest Number of Frames First}
\acrodef{LRC}{Network Computer Laboratory}

\acrodef{mMTC}{Massive Machine-Type Communications}
\acrodef{M2M}{Machine-to-Machine}
\acrodef{MAC}{Medium Access Control}
\acrodef{MCS}{Modulation and Coding Scheme}
\acrodef{MNO}{Mobile Network Operator}
\acrodef{MME}{Mobility Management Entity}
\acrodef{MMF}{Multi-mode Optical Fiber}
\acrodef{MPCP}{Multipoint Control Protocol}
\acrodef{MTP}{Multi-thread Polling}
\acrodef{MTU}{Maximum Transmission Unit}
\acrodef{MOS-IPACT}{IPACT with Multi-ONU with SLAs Support}
\acrodef{MPDCCH}{MTC \acs{PDCCH}}
\acrodef{MTC}{Machine-Type Communications}

\acrodef{nGBR}{non-GBR}
\acrodef{NGA}{Next Generation Access Network}
\acrodef{NO}{Network Operator}

\acrodef{OAM}{Operation Administration and Maintenance}
\acrodef{ODN}{Optical Distribution Network}
\acrodef{OLS}{ONU Load Status}
\acrodef{OMCI}{ONU Management Control Interface}
\acrodef{OFDM}{Orthogonal Frequency Division Multiplexing}
\acrodef{OFDMA}{Orthogonal Frequency Division Multiple Access}
\acrodef{ONU}{Optical Network Unit}
\acrodef{OPEX}{Operating Expenditure}

\acrodef{P2MP}{Point-to-Multipoint}
\acrodef{P2P}{Point-to-Point}
\acrodef{PDCCH}{Packet Downlink Control Channel}
\acrodef{PDSCH}{Physical Downlink Shared Channel}
\acrodef{PEDB}{Policies for Excess Bandwidth Distribution}
\acrodef{PHY}{Physical Layer}
\acrodef{PUSCH}{Packet Uplink Shared Channel}
\acrodef{PLOAM}{Physical Layer OAM}
\acrodef{PLR}{Packet Loss Ratio}
\acrodef{PRA}{Prioritized Random Access}
\acrodef{PRACH}{Physical Random Access Channel}
\acrodef{PDB}{Packet Delay Budget}
\acrodef{PUCCH}{Physical Uplink Control Channel}
\acrodef{PRB}{Physical Resource Block}
\acrodef{PCFICH}{Physical Control Format Indicator Channel}
\acrodef{PHICH}{Physical HARQ Indicator Channel}
\acrodef{P-GW}{Packet Data Network Gateway}
\acrodef{PER}{Packet Error Rate}
\acrodef{PS}{Packet Scheduling}
\acrodef{PRACH}{Physical Random Access Channel}
\acrodef{PDB}{Packet Delay Budget}
\acrodef{PON}{Passive Optical Network}
\acrodef{PPA}{Preamble-Priority Aware}
\acrodef{PPA+}[ePPA]{Enhanced Preamble-Priority-Aware}

\acrodef{QCI}{QoS Class Identifier}
\acrodef{QoS}{Quality of Service}
\acrodef{QPSK}{Quadrature Phase-Shift Keying}

\acrodef{RA}{Resource Allocation}
\acrodef{RACH}{Random Access Channel}
\acrodef{RAN}{Radio Access Network}
\acrodef{RAR}{Random Access Response}
\acrodef{ROI}{RAN Overload Indicator}
\acrodef{RRC}{Radio Resource Control}
\acrodef{RRS}{RACH Resource Separation}
\acrodef{RRM}{Radio Resource Management}
\acrodef{RAC}{Radio Admission Control}
\acrodef{RB}{Resource Block}
\acrodef{RLC}{Radio Link Control}
\acrodef{RAP}{Random Access Priorized}
\acrodef{RAO}{Random Access Opportunity}
\acrodef{RTT}{Round-Trip Time}

\acrodef{SAR}{Smallest Available Report First}
\acrodef{SR}{Scheduling Request}
\acrodef{SC-FDMA}{Single-Carrier Frequency Division Multiple Access}
\acrodef{S-GW}{Serving Gateway}
\acrodef{SDF}{Service Data Flow}
\acrodef{SRS}{Sounding Reference Signal}
\acrodef{SPS}{Semi-Persistent Scheduling}
\acrodef{SR}{Scheduling Request}
\acrodef{SLA}{Service Level Agreement}
\acrodef{SNS}{SubNetwork Slice}
\acrodef{SMF}{Single-Mode Optical Fiber}
\acrodef{SNMP}{Simple Network Management Protocol}
\acrodef{SPD}{Shortest Propagation Delay First}
\acrodef{STP}{Single Thread Polling}

\acrodef{T-CONT}{Transmission Container}
\acrodef{TA}{Time Alignment}
\acrodef{TC}{Transmission Convergence}
\acrodef{TCO}{Total Cost of Ownership}
\acrodef{TD}{Time-Domain}
\acrodef{TDD}{Time Division Duplexing}
\acrodef{TDM}{Time Division Multiplexing}
\acrodef{TDMA}{Time Division Multiple Access}
\acrodef{TDPS}{Time-domain Packet Scheduling}
\acrodef{THR}{Throughput}
\acrodef{TFT}{Traffic Flow Template}
\acrodef{TLBA}{Two-Layer Bandwidth Allocation}
\acrodef{TTI}{Transmission Time Interval}
\acrodef{TPC}{Transmit Power Control}
\acrodef{TSF}{Thread Scheduling Framework}
\acrodef{TWDM}{Time and Wavelength Division Multiplexing}
\acrodef{TWMD}{Time-Wavelength Division Multiplexing}
\acrodef{TWDMA}{Time and Wavelength Division Multiple Access}

\acrodef{UCCH}{Uplink Control Channel}
\acrodef{UE}{User Equipment}
\acrodef{UL}{Uplink}
\acrodef{US}{Upstream}
\acrodef{DS}{Downstream}
\acrodef{US}{Upstream}
\acrodef{UL-SCH}{Uplink Share Channel}
\acrodef{mIPACT}{modified-IPACT}

\acrodef{MSD}{Multiple-Scheduling Domain}
\acrodef{SSD}{Single-Scheduling Domain}
\acrodef{WA}{Wavelength Agile}

\acrodef{WF}{Water-Fill}
\acrodef{FF}{First-Fit}
\acrodef{6G}{6th Generation mobile networks}
\acrodef{AI}{Artificial Intelligence}

\acrodef{VoIP}{Voice-over-IP}
\acrodef{VNO}{Virtual Network Operator}

\acrodef{WDM}{Wavelength Division Multiplexing}
\acrodef{WDMA}{Wavelength Division Multiple Access}

\acrodef{X-GPON}{10 Gbit/s GPON}
\acrodef{XGEM}{GPON Encapsulation Mode}
\acrodef{XGPON}{10 Gigabit Capable PON}
\acrodef{XGTC}{XG-PON Transmission Convergence}

\acrodef{IXR}{Immersive eXtended Reality}

\acrodef{CS}{Central Server}
\acrodef{FSL}{Federated Supervised Learning}
\acrodef{FRL}{Federated Reinforcement Learning}

\acrodef{FedAvg}{Federated Averaging}
\acrodef{KPI}{Key Performance Indicator}

\acrodef{IoT}{Internet of Things}
\acrodef{RV}{random variable}
\acrodef{RAO}{random access opportunity}
\acrodef{CIoT}{cellular IoT}
\acrodef{LTE}{Long Term Evolution}
\acrodef{LTE-A}{LTE-Advanced}
\acrodef{LTE-M}{LTE-MTC}
\acrodef{NB-IoT}{Narrowband \ac{IoT}}
\acrodef{NR}{new radio}
\acrodef{mMTC}{massive machine-type communications}
\acrodef{MIMO}{multiple-input multiple output}
\acrodef{mMIMO}{massive \acl{MIMO}}
\acrodef{NOMA}{non-orthogonal multiple access}
\acrodef{MAC}{medium access control}
\acrodef{MAD}{maximum average distance}
\acrodef{ZC}{Zadoff–Chu}
\acrodef{FDD}{frequency division duplexing}
\acrodef{BS}{base station}
\acrodef{5G-NR}{5G new radio}
\acrodef{PRACH}{Physical Random Access Channel}
\acrodef{RACH}{Random Access Channel}    

\newcommand{\RevCarlos}[1]{{\color{black}#1\xspace}} 
\newenvironment{boldedcarlos}{\color{black}}{}

\newcommand{\U}{G\xspace}
\newcommand{\Li}{L\xspace}
\newcommand{\V}{V\xspace}
\newcommand{\F}{F\xspace}
\newcommand{\NS}{NS\xspace}
\newcommand{\NA}{NA\xspace}
\newcommand{\Inf}{U\xspace}
\newcommand{\Sim}{S\xspace}
\newcommand{\Ana}{A\xspace}
\newcommand{\Exp}{E\xspace}
\newcommand{\Yes}{Y\xspace}
\newcommand{\No}{N\xspace}
\newcommand{\BS}{BS\xspace}

\newcommand{\ONUDU}{ONU/DU\xspace}

\newcommand{\FF}{First-Fit\xspace}
\newcommand{\Forecast}{First-Fit-prediction\xspace}
\newcommand{\MOS}{MOS-IPACT\xspace}

\newcommand{\proposal}{proposal\xspace}

\newcommand{\issueone}{bandwidth request problem}
\newcommand{\issuetwo}{grant sizing policy problem}
\newcommand{\issuethree}{maximum cycle length problem}
\newcommand{\discipline}{policy\xspace}

\newcommand{\ie}{\emph{i.e.,}\xspace}
\newcommand{\eg}{\emph{e.g.,}\xspace}
\newcommand{\gformat}[1]{\mathcal{#1}}
\newcommand{\HSymbol}{H}
\newcommand{\E}{\gformat{E}}

\newcommand{\HGraph}[1]{\HSymbol(\V,\E)}
\newcommand{\Hypergraph}[2]{\HSymbol_{\code}(\V,\E)} 

\newcommand{\code}{\mathcal{C}\xspace}

    \DeclareMathSizes{10}{9}{7}{5}
    
    \newboolean{long}   

\begin{document}
\author{
\IEEEauthorblockN{Oscar J. Ciceri, Carlos~A.~Astudillo, Gustavo B. Figueiredo, Zuqing Zhu, and Nelson~L.~S.~da~Fonseca%
\thanks{O.~J.~Ciceri, C.~A.~Astudillo and N.~L.~S~da~Fonseca are with the Institute of Computing, University of Campinas 13083-852, Brazil (emails: oscar@lrc.ic.unicamp.br, castudillo@lrc.ic.unicamp.br, nfonseca@ic.unicamp.br).}%
\thanks{G.~B.~Figueiredo is with the Computer Science Department, Federal University of Bahia, Brazil (email: gustavo@dcc.ufba.br).}
\thanks{Z.~Zhu is with the School of Information Science and
Technology, University of Science and Technology of China, Hefei, Anhui 230027, P. R. China (email: zqzhu@ieee.org).}%
} }

\title{Passive Optical Networking for 5G and Beyond 5G Low-Latency Mobile Fronthauling Services}

\maketitle
\begin{abstract}
    Passive optical network (PON) technology offers an attractive cost-efficient alternative to support 5G and Beyond 5G mobile network fronthauling (MFH).
    However, MFH for such networks is challenging given its high bandwidth and strict latency requirements.
    To reduce these requirements, radio access network \RevCarlos{(RAN)} functional splitting has been introduced in 5G networks; this provides more flexibility in resource allocation since the protocol stack is distributed between the centralized and the distributed units. In contrast to the conventional MFH requirement of the RF-PHY 
    splitting, the MFH traffic produced by \RevCarlos{higher}-layer splittings becomes more dependent on the actual user traffic load.
    By capitalizing on the \RevCarlos{new} characteristics of the MFH traffic with \RevCarlos{RAN} functional splitting, this article introduces a \RevCarlos{resource allocation} mechanism to improve the performance of PONs serving MFH. 
\end{abstract}
\acresetall
\vspace{-2mm}
\IEEEpeerreviewmaketitle
\bstctlcite{IEEEexample:BSTcontrol}

\section{Introduction}
\label{sec:Introduction}

The \ac{CRAN} technology has recently been  proposed to deal with the high demands imposed by the need to provide services with stringent delay and high throughput requirements for  the \ac{5G} and \ac{B5G}. The employment of \ac{CRAN} saves energy by gathering shared resources into cloud-based facilities, such as the  Baseband Unit (BBU), which is responsible for processing signals sent by the Remote Radio Head (RRH) located in the cell site. On the other hand, the  physical separation of BBUs and RRHs imposes  a strong dependence on  \ac{MFH} capabilities.

The \ac{CPRI} is the most prominent interface and protocol adopted for \ac{MFH} \RevCarlos{in 4G networks}. 
However, depending on the base station (RRH) configuration, the use of \ac{CPRI} requires the availability of a huge fixed bandwidth capacity in the \ac{MFH}. 
The required MFH capacity can range from \unit[614]{Mbps}  to \unit[24.3]{Gbps}, with the delay and jitter required to synchronize the RRH with the BBU being \unit[250]{$\mu$s} and \unit[65]{ns}, respectively.
Such stringent requirements make impractical the adoption of CRAN for 5G and \ac{B5G} networks, since these can require MFH data rates as high as 
\RevCarlos{hundreds of  
gigabits per second}
due to the use of technologies such as \ac{mMIMO}, mmWave and terahertz communications. 

\begin{figure*}[!t]
    \centering
    \includegraphics[width=1.0\linewidth]{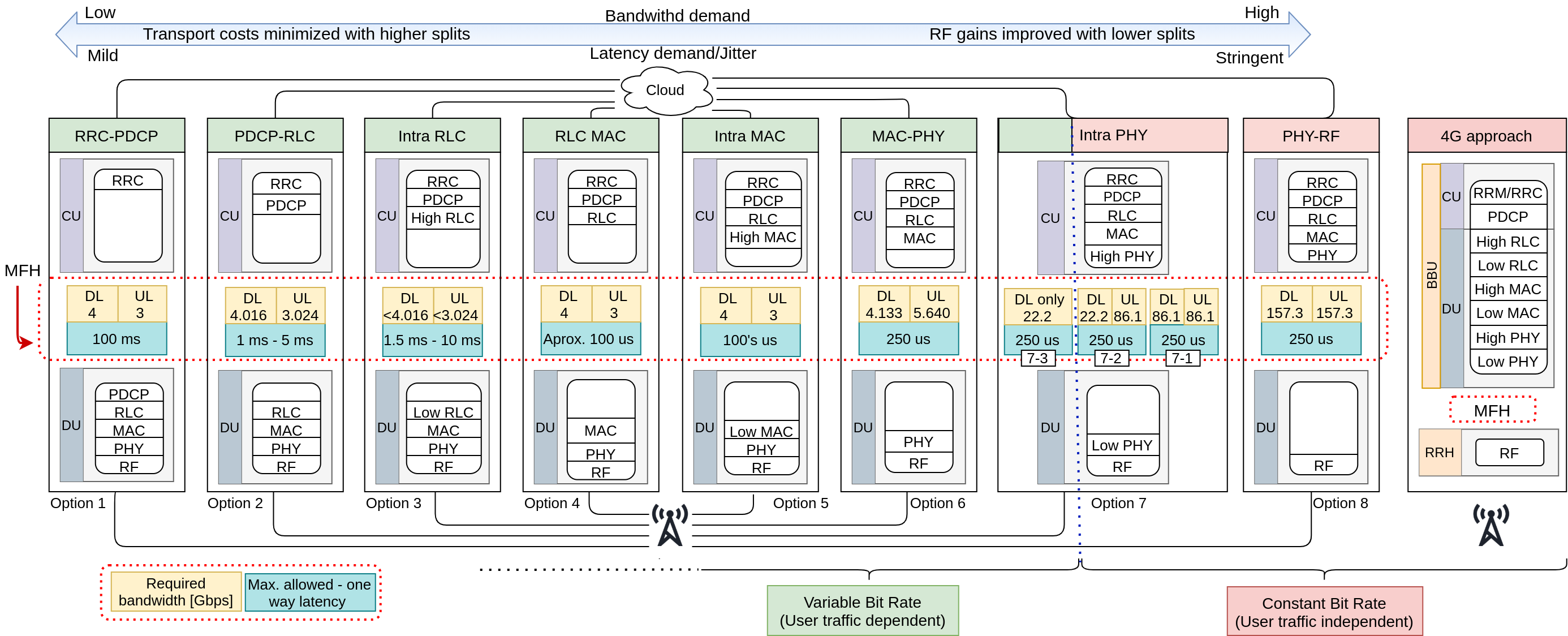}
     \caption{RAN functional split options. Delay requirements and bandwidth examples from 3GPP TR 38.801.}
    \label{fig:splitting_options}
\end{figure*}

To decrease the burden on the \ac{MFH}, the  3rd Generation Partnership Project (3GPP) has proposed the radio access network (RAN) functional splitting in Figure~\ref{fig:splitting_options}. 
It splits the functionality of BBUs by locating more RAN functions on the remote site and fewer on the centralized site than does the conventional BBU-RRH splitting of previous generations. This allows various possible \RevCarlos{split} options, \RevCarlos{which} 
provides flexibility and reduces the required bandwidth and delay in the fronthaul, although at the price of decreasing RAN centralization. 
\RevCarlos{Lower}-layer \RevCarlos{splittings} generate constant bit rates, whereas \RevCarlos{higher}-layer splittings imply MFH links with variable bit rates, which are dependent on the load of the user data plane. 

There is no imposition on the type of technology implemented in  \ac{MFH}, this can range from wireless to optical technologies. Due to its vast transmission capacity and current availability, the latter are preferable.  
\acp{PON} are attractive for MFH due to their point-to-multipoint topology, which is quite  suitable for fine granularity transport services.  
Moreover, recent efforts in PON standardization have resulted in a new architecture  specified in the 50G-EPON (\acs{IEEE}  802.3ca-2020), which employs the  \ac{TWDMA} technique for  upstream transmissions between  \acp{ONU} and  \ac{OLT}. It supports up to two \unit[25]{Gbps}  wavelength channels with non-tunable transceivers, increasing adequacy of the PON to support MFH traffic.

\ac{DWBA} algorithms for  resource allocation in PONs are typically based on on-demand schemes for granting bandwidth. 
In these schemes, clients request bandwidth in a scheduling cycle for transmission in the next cycle. This, however, adds queuing delays in the transmission of packets, which can be detrimental for the traffic of overloaded base stations. 
Resource allocation algorithms \RevCarlos{for the support of fronthauling traffic over PONs} that deal with load prediction, cooperation among \ac{5G} and \ac{PON} devices, 
and PON sharing have all recently been proposed (Table \ref{table:related_work}).  
\RevCarlos{However,} it is imperative that these \ac{DWBA} algorithms consider the tidal nature of mobile network traffic and capitalize on that \RevCarlos{spatiotemporal inbalance} to improve overall network performance and \RevCarlos{decrease MFH costs}.

This article addresses the employment of \ac{EPON} for mobile fronthauling in 5G and B5G networks  and proposes a \ac{DWBA} algorithm that  capitalizes on the opportunities emerging from the tidal nature of fronthaul traffic, as well as on the existence of customers leasing (owning) several ONUs from a single Infrastructure Service provider (multi-ONU customers), \RevCarlos{as in the case of the \acp{MNO}}. The proposed algorithm increases network utilization and statistical multiplexing gain for the \acp{MNO} that \RevCarlos{employ} 
the \ac{PON} for the support of \ac{MFH} traffic.

\begin{boldedcarlos}
\section{RAN Functional Splitting and its Implication to the Mobile Fronthualing Requirements}
\label{sec:Fronthauling}

Industry cooperation has recently taken place to decrease the \ac{MFH} data rate and delay requirements through RAN functional splitting, as illustrated in Fig. \ref{fig:splitting_options}. The CPRI protocol was also extended, resulting in a packetized protocol called eCPRI (enhanced CPRI), which supports these new split options.
The functional splits range from the simplest \ac{DU} (option 8) to the most complex \ac{DU} (option 1).

The conventional CPRI protocol used in 4G networks supports The PHY-RF option. In these networks, baseband processing is  centralized in the CU, and the DU  performs radio frequency (RF) functionality such as transmit and receive functions, filtering, amplification, analog-to-digital or digital-to-analog conversion, up/down conversion, and cyclic prefix (CP) removal/addition. 
This centralization of the RAN functions saves money by reducing the complexity of the remote units, the number of equipment, and energy consumption for running the network.
It also allows the sharing of the processing resources and facilitates cross-cell cooperation.
Advanced Coordinated Multipoint (CoMP) and soft handover schemes may be implemented at the CU to handle inter-cell interference and multi-cell transmissions. 
However, this split option requires a large amount of MFH capacity to transport in-phase and quadrature (I/Q) signal components between the remote and centralized locations.

The intra-PHY option has three variants. In option 7.1, the \ac{FFT} and \ac{iFFT} also reside in the DU, while the rest of the physical and link layer functions reside in the CU. 
Option 7.2 also adds  the pre-filtering/precoding and resource mapping/demapping to the DU. With option 7.3 (downlink only), the encoder also resides  at the CU. The fronthaul interface transport subcarriers, sub-frame symbols, and codewords when the options 7-1, 7-2, and 7-3 are employed, respectively.

The MAC-PHY option locates the Media Access Control (MAC) and upper-layer functionality at the CU, while the PHY and RF functions are located at the DU.
Thus, the MAC scheduler is centralized, and physical processing is performed locally in the DU. The transport blocks containing control and user data are transported through the MFH interface with this split option.

The intra-MAC option locates the real-time MAC functions with stringent delay requirements (\eg \ac{HARQ} and random access control) at the DU, whereas the less restrictive MAC functions such as resource scheduling, inter-cell interference coordination, and multiplexing functions are located at the CU. 
By locating the time-critical functions in the DU, the delay requirements on the fronthaul interface are relaxed  compared to the previous split options.

Centralized scheduling options (\ie Options 5, 6, 7, and 8) support CoMP functionality. Option 7-2 and lower-layer options fully support CoMP functions without performance degradation, while options 7-3, 6, and 5 restrict CoMP functionalities due to latency issues (\eg uplink joint reception). 

With the RLC-MAC option, the Radio Link Control RLC, Packet Data Convergence Protocol (PDCP), and Radio Resource Control (RRC) functions are located in the CU, while MAC and lower-layer functions reside at the DU. 
The tight coupling between RLC and MAC functions imposed the  most stringent latency requirements, especially for short subframes in 5G,requiring more frequent scheduler decisions.

The intra-RLC option splits the RLC layer into high-RLC and low-RLC located, respectively, at the CU and DU.
The segmentation process is performed in the low-RLC, while the \ac{ARQ} and other RLC functions reside in the high-LRC.
It is robust in non-ideal channel scenarios and scenarios with  high mobility.

In the PDCP-RLC option, the RRC and PDCP are located in the CU, while RLC, MAC, physical, and RF functions reside in the DU. 
This split option improves the traffic load control and traffic aggregation between the New Radio and Evolved Universal Terrestrial Radio Access transmission points.
Since the real-time MAC and RLC functions are performed locally in the DU, these last two options relax even more the MFH latency requirements. 

In the RRC-PDCP option, the CU handles the RRC function, while the rest of the RAN functions are located in the DU.
It separates the user-plane from the CU and improves traffic management in cases in which the user data needs to be placed near the transmission point (\eg edge computing and caching applications).

The low-layer related split options (7-1, 7-2, and 8) may generate very high constant bit rates in the MFH links, which depend on the cell bandwidth and the number of antennas, and  options 7-3 to 1 produce a variable bit rate MFH traffic which is more dependent on the actual user data plane traffic.
\end{boldedcarlos}

\section{Issues and Approaches for Supporting Mobile Fronthauling over EPON Networks}\label{sec:Issue}

Typically, \acp{InP} attempt to maximize network utilization and revenue by furnishing various services over the same \ac{PON} infrastructure to their customers, including \acp{MNO} and conventional customers such as residential users, enterprises and \aclp{VNO}.
Nevertheless, the unique requirements of MFH make \ac{QoS} provisioning a challenging task, especially in scenarios with coexisting MFH and conventional PON services.

\begin{table*}
\centering
\caption{Literature review on resource allocation for mobile fronthauling over PONs. \U: Gated; \Li: Limited; \F: Fixed; \V: Variable; \Inf: Unlimited; \Sim: Simulation; \Ana: Analytical; \Exp: Experimentation; \Yes: Yes; \No: No; \NS: Not Specified; \NA: Not Applicable}
\hspace*{-4mm}
\begin{tabular}{lrccccccccccccccc} 
\hline
\multicolumn{2}{c}{Feature} 
& This 
&\cite{Tashiro_6886953-OFC2014} 
&\cite{Bidkar_ICTON2020-9203123} 
&\cite{Zhou_JOCN2018-8273263} 
&\cite{Hiroshi_ICC2016-7511482} 
&\cite{Hisano_JSAC2018-8482302} 
&\cite{Nakayama_TCOMM2019-8823004} 
&\cite{Araujo_ICC2019-8761655} 
&\cite{Uzawa_JOCN2020-8896804} 
&\cite{Larsen_arxiv20201} 
&\cite{Hisano_JLT2017-7924359}
&\cite{Yuan_ACPC17}
&\cite{Zaouga_JLT02021-9293363}
\\ \hline\hline
\multirow{2}{*}{\textbf{Standard family}}  & 
IEEE (EPON)   & X & X & - & \multirow{2}{*}{\NS} & \multirow{2}{*}{\NS} & \multirow{2}{*}{\NS} & - & X & X & - & X & \multirow{2}{*}{\NS} & -\\ 
&   
ITU (GPON) & - & -  &  X  &  &   &  & X & - & - & X & - &  & X \\ 

\multirow{2}{*}{\textbf{Multi. access tech.}}  & TDMA-PON & - & X & X & X & X & X & - & - & X & X & X & - & -\\ 
&  TWDMA-PON  & X & - & - & - & - & - & X & X & - & - & - & X & X \\

\multirow{2}{*}{\textbf{Rate per wavelength}}  
& 10 Gbps  & - & X & X & X & X & X & X & X & X & X & X & - & - \\ 
& 25 Gbps & X & - & - & - & - & - & - & - & - & - & - & X & X \\

\multicolumn{2}{l}{\textbf{Wavelengths ($\lambda${s}) per OLT}} & 2  & 1 & 1 & 1 & 1 & 1 & 4 & \Inf & 1 & 1 & 1 & 16 & 2 \\
\multicolumn{2}{l}{\textbf{Simultaneous TX $\lambda${s} per ONU}}  & 1 & 1 & 1 & 1 & 1 & 1 & 1 & \Inf & 1 & 1 & 1 & 1 & 1 \\ 
\multicolumn{2}{l}{\textbf{Splitting option}} & 6 &  8 & \NS & 7.2 & 6 & 7.3
& 6 & 8 & 6 & 6 & 6 & 8 & 7.1\\
\multirow{2}{*}{\textbf{Traffic prediction}} &
BS-OLT coop. & X & X & \multirow{2}{*}{\No} & \multirow{2}{*}{\No} & X & - & \multirow{2}{*}{\No} & - & X & X & - & \multirow{2}{*}{\No} & \multirow{2}{*}{\No} \\ 
& Grant-Report & - & - &  &  & - & X &  & X & - & - & X &  &  \\
\multicolumn{2}{l}{\textbf{Maximum cycle length [ms]}} &
0.25 & 0.5 & \NA & \NS & V & 1 & \NA & \NS & 0.2 & \NA & 0.25 & \NS & \NA \\ 
\multicolumn{2}{l}{\textbf{SLA support}} & \Yes& \No &  \No &  \No & \No & \No &  \No &  \No &  \No & \No & \No & \No & \No \\ 
\multicolumn{2}{l}{\textbf{Conventional customers}} & \Yes & \No & \No & \No & \No & \No & \No & \No & \No & \No & \Yes & \No & \Yes \\ 
\multicolumn{2}{l}{\textbf{Grant sizing policy for MFH ONUs}} & \Li & \U & \F &  \F  & \U & L & \F & \U & \F & \U & \F & \NS & \U \\ 
\multicolumn{2}{l}{\textbf{Bandwidth Sharing}} & \Yes & \No &  \No &  \No & \No & \No &  \No &  \No &  \No & \No & \No & \Yes & \No \\ 
\multicolumn{2}{l}{\textbf{Perf. eval. based on real deployment.}}             
& \Yes & \No &  \No &  \No & \No & \No &  \No &  \No &  \No & \No & \No & \Yes & \No \\
\multicolumn{2}{l}{\textbf{3GPP TR-38.816 traffic modeling}} & \Yes &  \No & \No & \No & \No & \No & \No & \Yes & \No & \No & \No & \NS & \Yes \\
\multicolumn{2}{l}{\textbf{Publication year (20YY)}}
& 21 & 14 & 20 & 18 & 16 & 18 & 19 & 19 & 20 & 21 & 17 & 17 & 21 
\\\hline
\end{tabular}
\label{table:related_work}
\end{table*}

Various \RevCarlos{\ac{RA}} algorithms for EPON networks have been proposed for providing guaranteed bandwidth and low latency in MFH (\RevCarlos{See Table~\ref{table:related_work}}). In the following, the main issues of \RevCarlos{RA} algorithms and their approaches used to support of MFH in PONs are reviewed. 

The \RevCarlos{RA} algorithms can be classified into those for  \ac{SBA} and those for \acf{DBA}.
The former allocate a fixed transmission window for each ONU, independent of the ONU load, which guarantees deterministic delays, although bandwidth can be wasted, increasing  costs, especially when dealing with splittings with variable rates.
The latter allocate transmission windows on a per-cycle-basis depending on the offered load, delay requirement, and available bandwidth; they increase statistical multiplexing gain in scenarios with unbalanced loads but introduce challenges for the management of the available bandwidth in scenarios with low-latency requirements.  

\subsection{Problem of Bandwidth Requests}
\begin{figure}%
\centering
\subfloat[\titlecap{\issueone}]{\label{fig:report_problem}%
\includegraphics[height=2.1in]{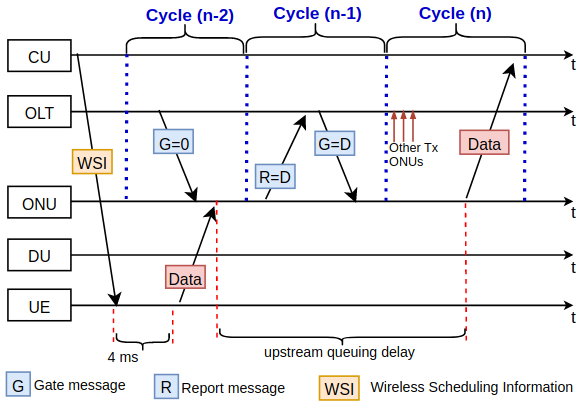}}%
\qquad
\subfloat[\titlecap{\issuethree}]{\label{fig:maximum_cycle_problem}%
\includegraphics[height=2.1in]{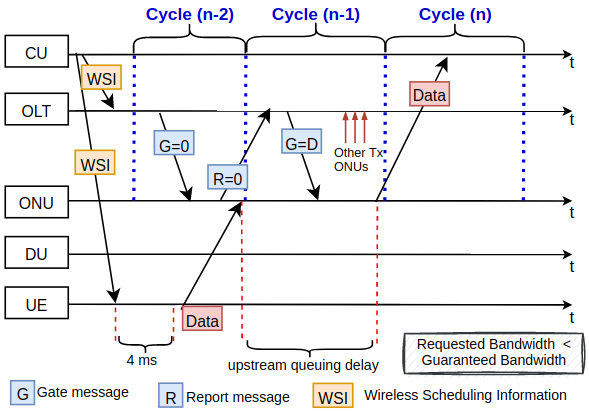}}%
\qquad
\subfloat[\titlecap{\issuetwo}]{\label{fig:grant_sizing_problem}%
\includegraphics[height=2.1in]{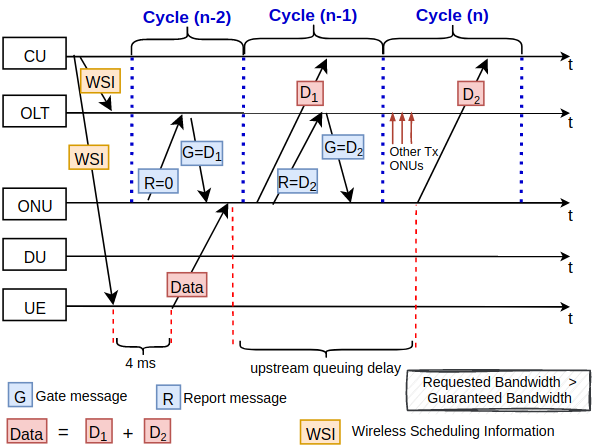}}%
\caption{Resource allocation issues for low-latency MFH over PONs}
\label{fig:issues}
\end{figure}

One of the most important issues  in \RevCarlos{RA} is the \textit{Problem of Bandwidth Requests} (Fig. \ref{fig:report_problem}). \ac{DBA} schemes use  Gate and Report messages
to coordinate upstream transmissions between the  \acp{ONU} and the \ac{OLT}. 
Bandwidth is requested in Report messages sent from the \acp{ONU} to the \ac{OLT} to request bandwidth,  whereas information on the bandwidth allocated in Gate message sent from the \ac{OLT} to the \ac{ONU}. Hence, the \ac{OLT} must wait for a request message before granting bandwidth, which implies that the upstream delay will be at least one scheduling cycle in duration. This delay can be as long as a millisecond, yet this is much longer than the required delay for split option 6 and above.

An approach to this problem involves estimations of accurate upcoming MFH traffic estimation  to avoid the need to wait for a report message, thus reducing  latency to acceptable levels. This approach use MFH traffic information (Wireless Scheduling Information (WSI) or traffic prediction based on report messages) to forecast  traffic arrivals in the near future and to allocate bandwidth without the need of  \ac{OLT} to receive a request, as shown in Fig. \ref{fig:maximum_cycle_problem}.

Cooperation between the CU/DU and the OLT was first proposed in \cite{Tashiro_6886953-OFC2014}, and has been widely adopted ever since as a key technique for low-latency MFH \RevCarlos{(\cite{Bidkar_ICTON2020-9203123, Hiroshi_ICC2016-7511482,
Araujo_ICC2019-8761655,
Uzawa_JOCN2020-8896804,Larsen_arxiv20201})}. This cooperation  allows the OLT to obtain accurate information about  upcoming traffic  by exploiting the WSI, which is used to inform mobile users about  resource allocation \unit[4]{ms} before the actual uplink transmission. 

\begin{figure*}[!t]
    \centering
    \includegraphics[width=0.9\linewidth]{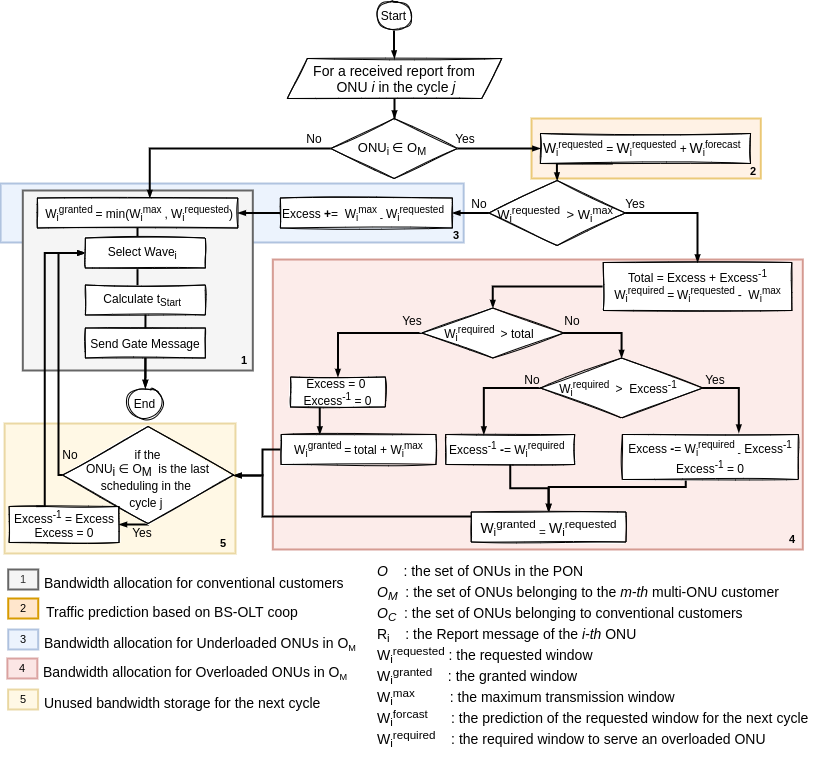}
     \caption{Flow chart of the proposed resource allocation algorithm}
    \label{fig:proposal}
\end{figure*}

\subsection{Problem of Maximum Cycle Length}
On the one hand, the upstream delay depends on the cycle duration because each \ac{ONU} usually transmits only  once per cycle.
If, for example,  mobile traffic arrives at the \ac{ONU} just after a transmission, these frames remain in the ONU queue at least until the next transmission cycle, as shown in Figure \ref{fig:maximum_cycle_problem}.
Even with MFH traffic estimation, this delay can be as long as the maximum cycle length when the network is  overloaded.
On the other hand, the polling overhead (or bandwidth wasted) mainly depends on the number of guard periods per cycle and the number of cycles per second. 
This means that the larger the maximum cycle length is, the lower the overhead. 
Thus, there is a trade-off between overhead and delay, which needs to be carefully addressed. As can be seen in Table \ref{table:related_work}, the importance of the maximum cycle length has not received the attention deserved in previous investigations \begin{boldedcarlos}(\eg \cite{Tashiro_6886953-OFC2014,
Zhou_JOCN2018-8273263,
Hiroshi_ICC2016-7511482,
Hisano_JSAC2018-8482302,
Araujo_ICC2019-8761655,
Yuan_ACPC17})\end{boldedcarlos}.

\subsection{Problem of Grant Sizing}
As seen in Table~\ref{table:related_work}, various resource allocation schemes for supporting MFH over PONs employ a Gated \discipline in which the \ac{OLT} allocates a transmission window equal to the requested/forecast one \begin{boldedcarlos}
(\eg \cite{Tashiro_6886953-OFC2014,Hiroshi_ICC2016-7511482,Araujo_ICC2019-8761655,Larsen_arxiv20201,Zaouga_JLT02021-9293363})\end{boldedcarlos}. 
However, this \discipline does not allow \ac{QoS} provisioning (\eg guaranteed bandwidth and delay) to diverse types of services coexisting on the same \ac{PON} infrastructure, as occurs with current shared \acp{PON}.
This problem occurs because a customer may overuse the total available bandwidth, thus increasing the cycle duration if no traffic shaping is undertaken.

\acp{InP} usually employ a limited type of \discipline  to guarantee bandwidth for  \ac{PON} customers according to a pre-defined \acp{SLA}. 
In this policy, the \ac{OLT} grants a transmission window equal to the minimum between the requested window and the maximum allowed transmission window (Fig. \ref{fig:grant_sizing_problem}). 
However, customers  who have  offered load greater than their guaranteed bandwidth in a scheduling cycle (overloaded customers) require various scheduling cycles to fully send their packets in the ONU queue.
The  uplink data queuing delay of these overloaded customers will depend on the number of cycles required for the OLT to provide the required bandwidth.
A bandwidth sharing mechanism has been proposed in \cite{ciceri2018dynamic} to address the bandwidth starvation problem of  overloaded \acp{ONU} in 4G backhauling scenarios with multi-ONU customers. 
This mechanism guarantees that the bandwidth of ONUs belonging to the same customer can be shared, so that unused bandwidth from underloaded \acp{ONU} can be assigned  to  overloaded ones on a per-cycle basis,  thus reducing the number of scheduling cycles needed to serve an overloaded customer.  
However, the calculation of the total unused bandwidth of an underloaded ONUs requires that the \ac{OLT}  wait for the arrival of all the Report messages from the ONUs belonging to that same customer before sending  Gate messages to the overloaded ONUs. This introduces latency. 

In summary, even though MFH traffic prediction can reduce the latency introduced by the \textit{\issueone}, the \textit{\issuetwo\xspace} and \textit{\issuethree\xspace} may still generate delays longer than those  required by the MFH service.

\section{DWBA Algorithm for EPON-based Mobile Fronthauling}
\label{sec:PON_standards}

To address the issues raised in the previous section, we introduce a novel \ac{DWBA} algorithm  to provide high-throughput and low-latency for 5G and B5G mobile fronthauling services, while meeting the requirements of the \acp{SLA} of all the PON customers.

The proposed mechanism adopts the widely-used cooperative approach proposed in \cite{Tashiro_6886953-OFC2014} to tackle the \textit{the problem of bandwidth requests}. 
To ameliorate the \textit{the problem of grant sizing} and exploit spatio termporal characteristics of MFH traffic, a bandwidth sharing mechanism is also employed.
The waiting time associated with bandwidth sharing techniques is avoided by employing an excess bandwidth compensation approach. 
The idea behind our proposal is to allow the use of the excess bandwidth of the previous and current cycles to serve the upcoming traffic of the overloaded \acp{ONU} of a multi-ONU customer in an online fashion.  
Moreover, the maximum cycle length is chosen so that it does not exceed the latency requirement, thus addressing the \textit{\issuethree}.
To the best of our knowledge, this is first solution to address \ac{QoS} provisioning for 5G and B5G MFH in shared \acp{PON}. 

Figure \ref{fig:proposal} summarizes the proposed scheme, which resides in the \ac{OLT}. 
When a Report message arrives from a conventional \ac{ONU}, the \ac{OLT} calculates the transmission window according to the limited policy (Block 1). 
If the Report message comes from a multi-ONU customer
, the \ac{OLT} updates the requested window by utilizing the optical-wireless cooperation procedure (Block 2). 
If the maximum window is greater than the requested windows, the ONU is fully served by the legacy transmission window and the unused bandwidth value is stored (Block 3). 
Otherwise, the \ac{OLT} grants additional bandwidth to the overloaded ONUs by utilizing the excess bandwidth from the previous and current cycles (Block 4).  
Moreover, when the scheduling cycle ends for the multi-ONU customer, the \ac{OLT} saves the remaining excess bandwidth value of the current cycle and discards the excess bandwidth of the previous one  (Block 5).
After this procedure, the \ac{OLT} allocates the upstream wavelength channel according to the First-Fit policy and calculates the next start time. The Gate message is then sent to the \ac{ONU}.

\section{Performance evaluation}
\label{sec:Performance_evaluation}
The performance of the proposed \ac{DWBA} scheme was evaluated  using an \ac{EPON} simulator (EPON-Sim), previously validated in \cite{ciceri2018dynamic}. EPON-Sim was extended to support TDWMA-PON and MFH service.  
We compare the performance of our proposal to that of \FF and \MOS \cite{ciceri2018dynamic} schemes both 
with and without the MFH traffic prediction technique. 

\subsection{Mobile Traffic Modelling}\label{sec:traffic_modelling}

A large dataset containing data of two \ac{MNO} cellular network deployments in Ireland \cite{di2017assembling} was used to capture the impact of relevant aspects such as topology, spatial traffic demands, and demographic and MNO information on the MFH performance.
This dataset provides base station location, operator, base station to user clusters association, area type of user clusters, and the \ac{CDF} of the demands of users served at the peak hour for each type of area.

Each \BS was classified as commercial, residential, or rural, on the basis of its most representative type of user served. 
\RevCarlos{Its} peak offered traffic load was generated by using Monte Carlo simulations.
\RevCarlos{First, we simulate the peak traffic load for each user served by employing the CDF of data demand for the corresponding type of area and aggregated the user cluster traffic demands corresponding to each base station.}

\begin{figure}[!t]
    \centering
    \includegraphics[width=0.8\linewidth]{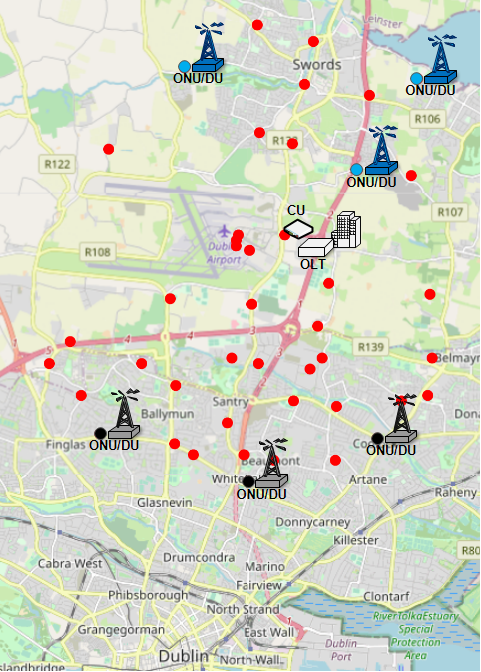} 
     \caption{Geographical location of base stations in a \unit[5]{km}-radius region in Dublin, Ireland. Points indicates base stations locations. Blue Points and black points indicate urban base stations (located in commercial areas) and suburban base station (located in residential areas) used in the simulations, respectively.}
    \label{fig:map_base_stations}
    \centering
\end{figure}

An \ac{InP} with an \ac{EPON} system with a 5 km radius coverage to the  North of Dublin was assumed.
In that region, one of the \acp{MNO} owns 44 \BS{s}, out of which 40\% serve residential areas; the rest serve commercial areas. Moreover,  about 15\% of the base stations (six) was assumed to be employing that \ac{PON} as their MFH.
The \BS traffic distribution was assumed to vary depending on \BS location and time period 
following \cite{6897914} so that the tidal effect can be fully captured and exploited in the performance evaluation. 
Thus,  three commercial and three residential base stations within the  region were selected (shown in Fig. \ref{fig:map_base_stations}) such that the Jain fairness index of  offered load in peak hours was maximized in the selected region. The \BS{s} shared their guaranteed bandwidths when bandwidth sharing based schemes were employed. 

In this way,  the \BS offered loads for each scenario were obtained, as well as the distances between the OLT and the MFH ONUs used in the simulations. The obtained load values were scaled with a factor $c=3$ to cope with  5G demands.
The scaled offered loads obtained were \unit[(25.51; 30.09; 21.46)]{Mbps} and \unit[(27.45; 23.36; 30.00)]{Mbps} for the residential and commercial \BS{s}, respectively.

\subsection{Simulation Model and Setup}
\label{sec:simulation_model}

A 50G \acs{TWDMA}-EPON network with a tree topology and an \ac{OLT} that handles a set of $32$ \acp{ONU} was simulated. Each ONU can transmit on a single 25G wavelength that is allocated dynamically.
There is a \ac{MNO} employing the MFH service of an InP  to serve six \BS{s}, as described in Section \ref{sec:traffic_modelling}. 
Each MFH \ac{ONU} is connected to its corresponding DU through a local 100Gbps Ethernet interface. 

We assumed the split option 6 for all the \BS{s} because it is the most demanding one with the largest bandwidth and lowest latency requirements among the \RevCarlos{upstream} variable-rate split options (see Fig. \ref{fig:splitting_options}).
The \BS peak loads obtained in Section \ref{sec:traffic_modelling} and the same assumptions proposed in 3GPP TR 38.801v14 were used to generate the peak average offered load of the $k$th \ONUDU ($P_k$).
The obtained $P_k$ values were
\unit[(4170, 4445 and, 3927)]{Mbps}
and
\unit[(4287, 4041, and 4440)]{Mbps}, respectively, for the residential and the commercial \ONUDU{s}.

The guaranteed bandwidth $B_k$ of the MFH \acp{ONU} was varied from $0.8\cdot P_k$ to $1.2\cdot P_k$. For the sake of clearness, herein after, $P_k$ has been omitted from the $B_k$ values.
To test a coexisting scenario with support for different PON services, the rest of the \acp{ONU} in the PON were conventional ones.
Each conventional ONU had a guaranteed bandwidth equal to the remaining available bandwidth in the \ac{PON} divided by the number of conventional \acp{ONU}. Moreover, to simulate a highly loaded network condition, the mean offered load of the conventional \acp{ONU} was \unit[85]{\%} of their guaranteed bandwidth.
\RevCarlos{It was pointed out by a Brazilian InP that, in practice, additional network resources are allocated to the system when it achieves around \unit[85]{\%} of its capacity.}

The load generated by the ONU/DU follows a Poisson distribution with a mean value equal to the offered load for the corresponding scenario, while that for the traffic of conventional \acp{ONU} follows the implementation in \cite{ciceri2018dynamic}.
Moreover, DU was assumed to generate bursts of Ethernet frames (MFH data) every \unit[250]{$\mu$s}.   
The maximum cycle length was set to \unit[250]{$\mu$s}, the propagation delay in fiber was considered to be \unit[5]{$\mu$s/km}, and the guard time period between transmissions from different \acp{ONU} in the PON was \unit[0.624]{$\mu s$}. 
Each simulation scenario lasted \unit[60]{s} and was replicated $10$ times.

Two different time period scenarios were considered, namely 18h and 24h, as these account for the peak of the offered traffic loads for commercial and residential \BS{s}, respectively.
\begin{boldedcarlos}The offered load for residential \BS{s} on 18h and commercial \BS{s} on 24h were, respectively, 38.1\% and 8.1\% of that during the peak hour.\end{boldedcarlos}
Since MFH ONUs serving \BS{s} in off-peak traffic hours experience lower delay values than do those \begin{boldedcarlos}serving \BS{s} in\end{boldedcarlos} peak traffic hours,
the analysis in the next section focuses on MFH ONUs in the peak traffic hour  for each time scenario.

\subsection{Simulation Results and Discussions}\label{sec:results}

\begin{figure*}%
\centering
\subfloat[Commercial area \ONUDU{s} at 18h]{\label{fig:r3}
\includegraphics[height=2.2in]{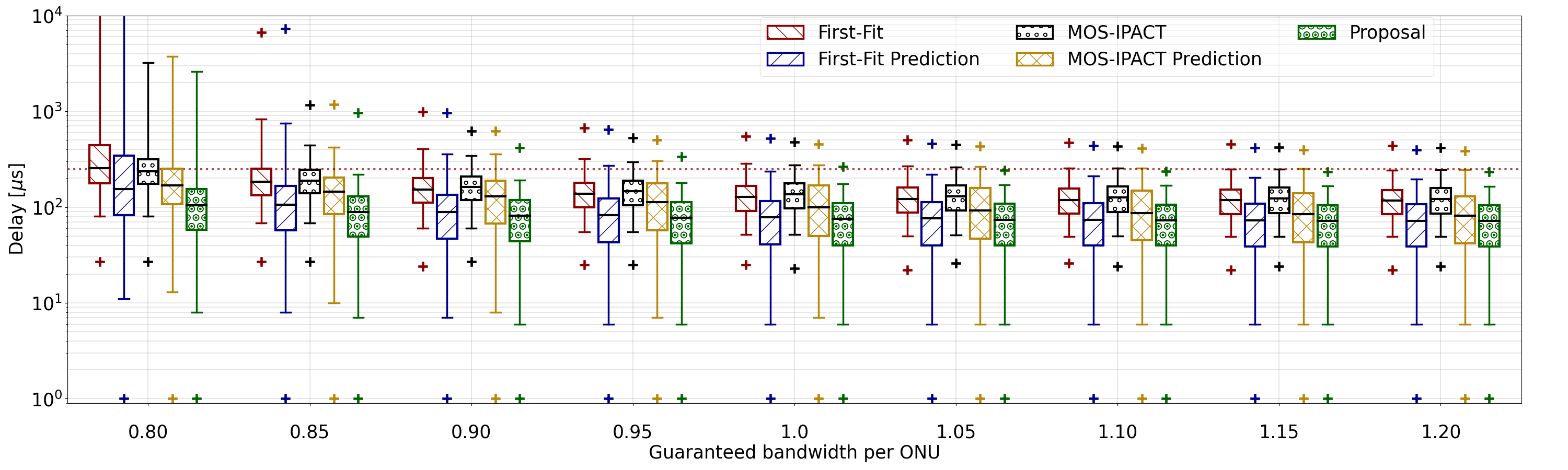}}%
\qquad
\subfloat[Residential area \ONUDU{s} at 24h]{\label{fig:r4}
\includegraphics[height=2.2in]{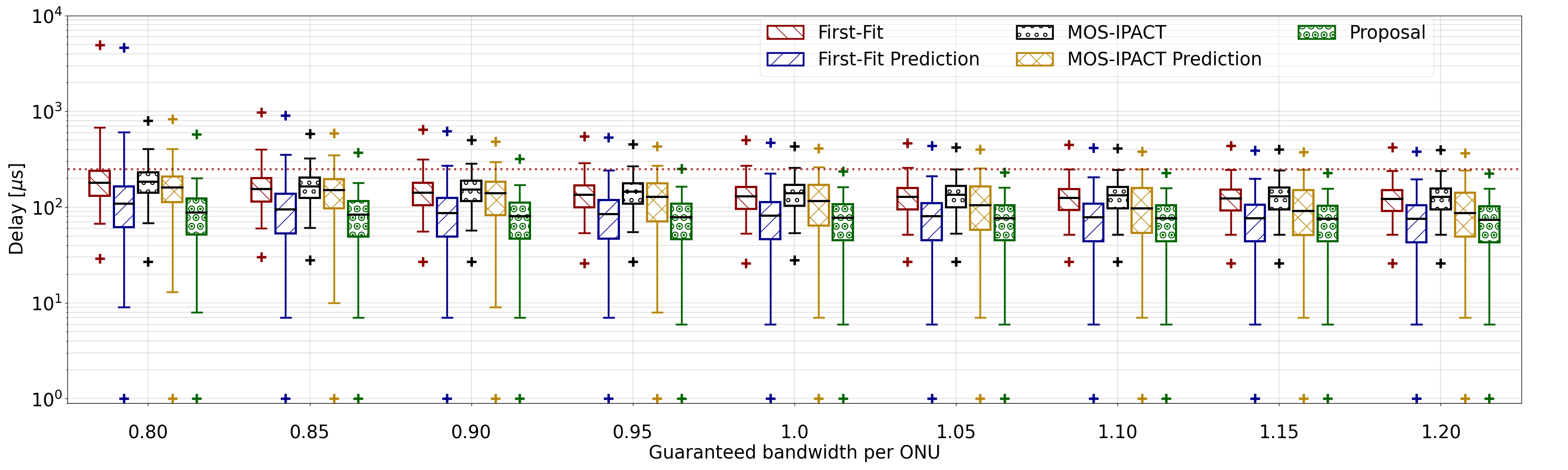}}%
\caption{
Delay performance of MFH ONUs for different resource allocation scheme with RAN \RevCarlos{functional split} option 6; Each boxplot shows  1st, 25th (Q1), 50th (Q2), 75th (Q3), and 99th percentile of delay values;
the lowest and the 99.999th percentile delay values are shown as lower and upper outlier points, respectively.
}
\label{fig:results}
\end{figure*}

The proposed scheme achieved the required delay value (<250$\mu$s) for both  scenarios 
with guaranteed bandwidth per \ONUDU greater than or equal to 105\% of its peak hour average load value (Fig. \ref{fig:results}). Note that the MFH ONUs in the 18h scenario require 10\% more guaranteed bandwidth than do those in the 24h scenario because of the differences in the BS off-peak traffic loads of the two scenarios. Since the bandwidth leased has to satisfy the worst-case scenario, the MNO needs 1.05 guaranteed bandwidth per \ONUDU. 
The other schemes fail to produce satisfactory MFH delay values for the splitting option considered for the guaranteed bandwidth values tested.

These results show not only that the MFH delay requirements can be met by our proposal but also that 
it significantly reduces the required guaranteed bandwidth per \ONUDU when compared to the other schemes, thus leading to an increase in bandwidth utilization and a reduction in costs for \acp{MNO} renting bandwidth from the \ac{InP}.

The schemes employing MFH traffic prediction reduce the percentiles of delay values when compared to their version with no  prediction scheme. 
Moreover, bandwidth sharing based schemes give 99.999th percentile delay values lower than those without this technique.
Furthermore, online-based with traffic prediction schemes (\ie \Forecast, and our \proposal) generate 99th percentile delay values lower than those using offline policies. 
Our proposal combines all these features to meet the stringent MFH requirements of a network which provides QoS guarantees for all supported services.

\RevCarlos{
The evaluated schemes do not decrease the  99.999th percentile of the delay value after a particular  increase in the contracted bandwidth. There is a fundamental limit in the performance of  resource allocation mechanisms that cannot be surpassed by simply increasing the bandwidth. 
Even with accurate MFH traffic forecasting and appropriate maximum cycle length settings, the \emph{grant sizing problem} must be properly addressed to meet MFH low latency  requirements.}
 
\begin{boldedcarlos}
\section{Future Research Directions}
\label{sec:future_directions}

This section presents some future research directions for the application of \ac{PON} technology in fronthauling 5G and B5G networks.

\subsection{MFH Traffic Forecasting}
Most of the current approaches for forecasting MFH traffic employ information from the cooperation interface between the BS and the OLT. However, for splitting 5 and below, the direct cooperation between the OLT and the 5G MAC layer is not feasible since the required information is available only at the \ONUDU side.
Alternatively, the \ONUDU and the OLT could communicate to provide the information needed for traffic forecasting.
However, such communication requires additional bandwidth and introduces latency, which can only be afforded by \ac{eMBB} and \ac{mMTC} services.

Another option is to perform forecasting at the \ac{ONU} 
either with or without ONU-DU cooperation and include the predicted value in conventional report messages. 
There is a trade-off between latency overhead and prediction accuracy. Such trade-off is especially relevant when involving \ac{URLLC} since it may require a one-way access delay as low as \unit[100]{$\mu{s}$}.
Machine learning algorithms can be employed for traffic prediction with high accuracy and in acceptable time frames to cope with this trade-off.

\subsection{MFH-Aware PON Dimensioning and Planning}
The designs of PONs have not considered the requirements of MFH traffic.  Principles in PON design need to be defined to maximize network utilization yet minimizing the guaranteed bandwidth of a group of ONUs and supporting the demands under various delay constraints.
For instance, groups of \ONUDU{s} could be employed to optimally exploit the bandwidth sharing concept. Another potential approach is the design of an MFH network to reduce the number of wavelengths and OLT equipment required.

\subsection{MFH Traffic Management}
Traffic management at the ONU plays a key role in  supporting  low-latency MFH traffic over PONs, especially for variable-rate splitting options under dynamic resource allocation schemes that employ the Gated bandwidth allocation policy. 
Traffic management mechanisms are essential to guarantee deterministic delay bounds. 
For instance, the value of the parameters of traffic shaping mechanisms should be optimal for the delay requirement of each
splitting option. Such tuning needs to be well understood to achieve compliance with the definition of splitting options.
\end{boldedcarlos}

\section{Conclusion}
\label{sec:Conclusion}
This paper has introduced a novel resource allocation (RA) mechanism for supporting 5G and B5G mobile network fronthauling (MFH) in \acp{EPON}.  The main RA issues and approaches for supporting low-latency MFH in these networks are discussed.
Our proposal includes bandwidth sharing for multi-ONU customer, MFH traffic prediction and maximum cycle length tailored to MFH traffic requirements.
Simulation results show that our proposal provides lower delay values than do existing schemes under realistic traffic scenarios.
The proposal increases network utilization and statistical multiplexing gain for an \ac{MNO} employing \ac{PON}-based \ac{MFH} services. This leads to  lower MFH costs than those  of  existing approaches.
Moreover, with our proposal, InPs can offer attractive business models for MFH services.

\section*{Acknowledgment}
The authors would like to thank CNPq and the Sao Paulo Research Foundation (FAPESP) for grant 2015/24494-8.

\bibliographystyle{IEEEtran}
\bibliography{PON_for_5G_Fronthauling}

\section*{Biographies}
\begin{IEEEbiographynophoto}
{Oscar J. Ciceri} 
completed his five-year degree in Electronics and Telecommunications Engineering at Universidad del Cauca (UNICAUCA),  Colombia,  in 2015, and his M.S. degree in Computer Science at the University of Campinas (UNICAMP),  Brazil, in 2019. He is a Ph.D. student and researcher in the Network Computing Laboratory (LRC) at that university. He has received two IEEE best conference paper awards. His research interests include Passive Optical Networks (PONs), 5G and 5G beyond networks, Machine Learning, and virtualization. His current work is on the low-latency PON technologies for 5G and 5G beyond mobile optical access systems.%
\end{IEEEbiographynophoto}
\begin{IEEEbiographynophoto}
{Carlos A. Astudillo}
received his BEng degree in Electronics and Telecommunications Engineering from the University of Cauca in 2009, and his MSc degree in Computer Science from the University of Campinas (UNICAMP) in 2015. Currently, he is a PhD candidate at the Institute of Computing (IC), UNICAMP. He received the best thesis award 
at the 29th theses and dissertations contest of the Brazilian Computer Society, the 2016 best master’s thesis award from IC/UNICAMP, and two IEEE best conference paper awards. His current research includes resource allocation in 5G \& B5G cellular networks and machine learning for communications and networking.%
%
\end{IEEEbiographynophoto}
\begin{IEEEbiographynophoto}
{Gustavo Bittencourt Figueiredo} [M'14, SM'18] received his B.Sc. degree in Computer Science from Salvador University (2001), and the M.Sc. (2003) and Ph.D. (2009) degrees in Computer Science from the University of Campinas. Since 2010, he has been affiliated with the Department of Computer Science at the Federal University of Bahia, Bahia - Brazil, where he is currently an Associate Professor. His main research interest includes Optical Networks, 5G Networks, Networking Modeling and Optimization, and the Design of algorithms for networking.%
\end{IEEEbiographynophoto}
\begin{IEEEbiographynophoto}
{Zuqing Zhu} [M’07, SM’12] 
received his Ph.D. degree from the Department of Electrical and Computer Engineering, University of California, Davis, in 2007. He is currently a full professor at USTC. He has published more than 200 papers in referreed journals and conferences. He was an IEEE Communications Society Distinguished Lecturer (2018–2019) and a Senior Member of OSA.
\end{IEEEbiographynophoto}
\begin{IEEEbiographynophoto}
{Nelson L. S. da Fonseca} [M'88, SM'01] 
obtained his Ph.D. degree from the University of Southern California in 1994. He is Full Professor at the Institute of Computing, State University of Campinas (UNICAMP). He published 400+ papers and supervised 70+ graduate students. He is Senior Editor for the IEEE Communications Magazine and the IEEE Systems Journal. He is also an Associate Editor for Peer-to-Peer Networking and Applications and Computer Networks. He is past EiC of the IEEE Communications Surveys and Tutorials. He served as the IEEE ComSoc VP Publications, VP Technical and Educational Activities, and VP Member Relations.%
\end{IEEEbiographynophoto}
\end{document}